**Determination of Thermal History by Photoluminescence of Core-shelled Quantum Dots Going Through Heating Events**

*Yucheng Lan\*, Hui Wang, Nitin Skula, Xiaoyuan Chen, Yalin Lu, Gang Chen, and Zhifeng Ren\**

Prof. Yucheng Lan
Department of Physics and Texas Center for Superconductivity, University of Houston, Houston, TX 77204, USA
Department of Physics and Engineering Physics, Morgan State University, Baltimore, MD 21251, USA
E-mail: yucheng.lan@morgan.edu

Mr. Hui Wang, Prof. Zhifeng Ren
Department of Physics and Texas Center for Superconductivity, University of Houston, Houston, TX 77204, USA
E-mail: zren@uh.edu

Dr. Nitin Skula, Dr. Xiaoyuan Chen, Prof. Gang Chen
Department of Mechanical Engineering, Massachusetts Institute of Technology, Cambridge, MA 02139, USA

Prof. Yalin Lu
Laser Optics Research Center, US Air Force Academy, Colorado Springs, CO 80840, USA

A kind of novel thermal history nanosensors were theoretically designed and experimentally demonstrated to permanently record thermal events. The photoluminescence spectrum of core-shelled quantum dots CdSe/ZnS irreversibly shifted with heating histories (temperature and duration) of thermal events. The induced photoluminescence shift of the quantum dots CdSe/ZnS was employed to permanently record thermal histories. We further modeled a kind of thermal history nanosensor based on the thermal induced phenomena of core-shelled quantum dots to permanently record thermal histories at microscale and demonstrated to reconstruct temperature and duration of heating events simultaneously from photoluminescence spectra of the quantum dots. The physical mechanism of the sensors was discussed.

## 1. Introduction

Thermometry concerns with the design of thermal sensors and the measurement of temperature, playing important roles in scientific activities, industrial applications, and human lives to measure real-time temperature and permanently record thermal history.[1] Macroscopic thermal history sensors, such as proxy thermometers and register thermometers, can permanently record temperatures that can be read out later, much different from real-time thermal sensors that only real-time measure temperatures. For example, tree rings and ice cores can naturally record climate temperatures up to two thousand years [2] and 420,000 years [3] respectively, and the recorded prehistoric temperatures can be read out now from the ratio of oxygen isotopes by scientists. Nanoscaled thermal history sensors, nanoscaled counterparts of the macroscopic thermal history sensors, have been developed recently to permanently record temperatures of heating events. [4-7] These thermal history nanosensors utilized irreversible physical properties of nanoparticles changing with temperatures to record temperatures permanently, much different from real-time thermal nanosensors [8-18] in which physical changes are reversible with temperatures. Up to now several techniques have been developed to read out permanently recorded thermal histories, such as electron microscopy, [4,5] surface plasmon absorption spectroscopy [7] and Raman spectroscopy [6]. The recorded temperatures were read out after heating events from location of nano-oxides,[4] the size of quantum dots,[5] Raman spectrum,[6] and surface plasmon [7] *etc*. However, these approaches could only read out the permanently recorded highest temperatures of thermal events [4,5,7] at nanoscale, or average heating temperatures and duration of heating events [6] at microscale. There are no thermal history sensors existing to permanently record heating temperature and heating duration simultaneously at microscale.

In the present work, core-shelled quantum dots (QDs) were utilized as thermal history nanosensors to permanently record both temperature and duration of thermal events at microscale. The thermal histories (temperature and duration) were reconstructed later from irreversible optical properties of the quantum dots, providing fingerprints for thermal history data retrieving. The demonstrated QD-based thermal history sensors are potentially applicable to many environments where the real-time temperature monitoring is unapproachable while a thermal history should be auto-recorded at microscale.

## 2. Results and Discussion

### 2.1 Design and Mechanism of Core-shelled QD-Based Thermal History Sensors

Semiconducting core-shelled quantum dots are excellent candidates to permanently record thermal histories that can be read out from optical spectrum. **Figure 1a** illustrates the energy band structure of core-shelled CdSe/ZnS quantum dots. The quantum confinement of the CdSe cores depends strongly on its size and the surrounding energy barrier. When a temperature field is applied to the quantum dots, the heating procedure would cause mass diffusions across the core and shell interfaces of the quantum dots, changing the energy structure of the QDs irreversibly (Figure 1b) due to quantum confinement effects. The changed energy structure thus can permanently record the thermal history. The change of energy structure would affect optical properties of the QDs. Subsequent to the heating event, the change of energy structure can be measured from photoluminescence spectrum and then the temperature history can be extracted from the quantum confinement of the heated QDs.

The band gap energy $E_{gap}$ of CdSe cores with a radius $R$ can be approximated as [19]

$$E_{gap} = E_{bulk} + \left[ \pi^2 \left( \frac{a_B}{R} \right)^2 - 1.786 \left( \frac{a_B}{R} \right) \right] R_y^* \quad (1)$$

where $E_{bulk}$ is the band gap of CdSe bulks. $E_{bulk}$ = 1.74 eV. The second term of **Equation 1** comes from the quantum confinement effect and exciton energy of the CdSe QD cores. $a_B$ is the Bohr radius of CdSe cores ($a_B$ = 4.9 nm) and $R_y^*$ the effective Rydbergy's energy. $R_y^* = -\frac{1}{\varepsilon^2} \frac{\mu}{m_e^*} R_y$, where $R_y$ is the Rydberg's energy with $R_y$ = 0.016 eV, $\mu$ the reduced mass ($1/\mu = 1/m_e^* + 1/m_h^*$, $m_e^*$ = 0.13 $m_e$ = 1.18 × 10$^{-31}$ kg and $m_h^*$ = 0.45 $m_e$ = 4.09 × 10$^{-31}$ kg are the excited electron mass and hole effective mass of CdSe, respectively), and $\varepsilon$ the dielectric constant.

After thermally exposed to a thermal event with heating temperature $T$ and heating time $t$, the thermal mass diffusion at the interface of the CdSe cores and ZnS shells would reduce the CdSe core size $R$ and change shapes of its energy well and the ZnS barriers, which is illustrated in **Figure 1b**. The mass diffusion thickness is proportional to $\sqrt{D}t^n$ where the mass diffusivity $D = D_0 e^{-\frac{E_b}{\kappa_B T}}$, $E_b$ is the energy barrier of the mass diffusion, $\kappa_B$ the Boltzmann constant, and $n$ the time exponent. The band gap energy $E_{gap}$ of heated CdSe cores would shift with the heating temperature $T$ and time $t$ because of the thermal diffusion and quantum confinement of the QD cores. The energy shift $\Delta E_{gap}(T,t)$ can be expressed as

$$\left| \Delta E_{gap}(T,t) \right| = L_0(R)\sqrt{D_0}\, e^{-\frac{E_b}{2\kappa_B T}} t^n$$

$$L_0(R) = \left[ 2\pi^2 \left( \frac{a_B^2}{R^3} \right) - 1.786 \left( \frac{a_B}{R^2} \right) \right] R_y^* \qquad (2)$$

by taking derivative of Equation 1 with respect to radius $R$ and applying typical solution of the diffusion equation for the diffusion length, depending on the heating temperature $T$ strongly but weakly on the heating time $t$. **Equation 2** indicates that the energy shift induced by mass diffusion in the QDs follows Arrhenius-type kinetic process, and is very sensitive to the core size $R$.

The thermal induced energy shift can be measured from photoluminescence (PL) spectroscopy. The energy barrier $E_b$, the diffusion parameter $D_0$, and the time component $n$ in Equation 2 can be experimentally determined by fitting PL data. Then the permanently recorded temperature $T$ and duration $t$ can be reconstructed experimentally. The concept of QDs-based thermal history sensors will be demonstrated in Results and Discussion.

In the demonstration, PL spectra of CdSe/ZnS quantum dots were first characterized at various heating temperature $T$ and heating time $t$. Several kinds of CdSe/ZnS core-shelled QDs were examined and their physical parameters, $D_0$, $E_b$, and $n$ were calibrated. Then these kinds of QDs were employed to permanently record thermal histories. A recorded thermal history (heating temperature $T$ and duration $t$) was read out from PL spectra of the heated QDs as a demonstration.

## 2.2. Thermal Dependent Photoluminescence of CdSe/ZnS QDs

The photoluminescence spectrum of CdSe/ZnS core-shelled QDs strongly depends on heating histories. **Figure 2**a illustrates PL spectra of the QDs heated at various temperatures $T$ for 16 s. PL signals were readily detectable even when the heating temperature $T$ was up to 500 $^o$C. The PL peak shifted with heating temperatures $T$.

In order to quantitatively investigate the relationship between PL peak and heating temperatures $T$, the PL spectra shown in Figure 2a were fitted using the Gaussian function and the wavelengths of the peaks were calculated from the fitting. Most PL spectra were symmetric and fitted well using a Gaussian function. Some PL spectra, especially these from heated QDs over 450 $^o$C, were nonsymmetric. For these nonsymmetric spectra, a Gaussian function was used to fit the strongest peak. Figure 2b shows the temperature dependence of PL peak wavelength of the heated QDs. The PL peak wavelength changed very little with heating temperatures $T$ when $T < 150$ $^o$C while exhibited a significant blue shifting with heating temperatures when $T = 150 - 500$ $^o$C. In the high temperature range, the PL peak wavelength decreased sharply above 300 $^o$C. The PL peak shift provided a fingerprint of irreversible optical properties of the QDs during the thermal exposures the QDs experienced.

The QDs were also heated in air for other heating time $t$ ($t = 0.5 - 128$ s) at high temperatures. All PL spectra of QDs were collected at room temperature after the QDs were heated for 0.5 - 128 seconds at a certain temperature between 150 $^o$C and 500 $^o$C. No PL signal was detectable after the QDs were fast heated over 500 $^o$C in air. During $T = 150 - 500$ $^o$C, similar PL spectrum and temperature dependent PL peak shown in Figure 2 were observed. The PL peak blue-shifted with the heating temperature $T$ in all of these cases for a certain heating time $t$. Below 150 $^o$C, the PL peak shifts were too small to be detected.

The PL peak wavelength also depends on the heating time $t$ besides heating temperature $T$. **Figure 3**a illustrates typical PL spectra of the QDs that were heated for a series of time $t$ at 500 $^o$C. The PL peak wavelength blue-shifted with the heating time $t$. Figure 3b shows the heating time dependent optical behavior of the QDs heated at 500 $^o$C. The PL peak shifted to higher energy with the heating time $t$. The PL peak wavelength was sensitive to heating and shifted with heating time $t$.

The blue-shift of the PL peak with heating time $t$ was also observed at other heating temperatures $T$ from 150 $^o$C to 500 $^o$C. The obtained PL spectra and time dependence of PL peak were similar to these shown in Figure 3.

PL spectra of the CdSe/ZnS QDs strongly depends on the heating temperature $T$ but weakly on heating time $t$. Figure S2 illustrates the PL peak-$T$-$t$ relationship. For a thermal event, the heating temperature $T$ can be extracted from PL peak shifts when the heating time $t$ is known or the heating time $t$ can be extracted when the heating temperature $T$ is known, from the PL peak-$T$-$t$ relationship shown in Figure S2. Therefore, by characterizing time-temperature response of the QDs during heating processes, the history of a thermal event can be mapped out in a targeted sensing range of both temperature and duration. Figure S3 demonstrated how a temperature distribution was permanently recorded and read out along a hotwire with a diameter of 25 microns using the first kind of QDs.

At the meantime, the PL peak intensity also decays with heating temperature $T$ due to the defect creation that quenches the luminescence. In practical applications of the thermal history nanosensors, the amount of QDs, which directly affect the PL peak intensity, is hard to control. Therefore, in the following discussion, we only employed PL peak wavelength to detect a thermal event and do not use PL peak intensity.

The transmission electron microscopy images indicated that the size of the core-shelled QDs changed very slightly (Figure S4) during the fast heating. The origin of the PL peak shift with heating should be mainly contributed to the mass diffusion, such as the diffusion of ZnS shells into CdSe cores and CdSe cores into ZnS shells (Figure 1). More detailed mass diffusion is being carried out at nanoscale on a scanning transmission electron microscope. The mass diffusion should reduce the size of the pure CdSe cores and induce stronger quantum effects to broaden the band-gap of CdSe/ZnS QDs.

### 2.3. Thermal Dependent Band-gap of CdSe/ZnS QDs

The relationship between PL peak shift and heating temperature $T$, heating time $t$ can be expressed by Equation 2. In Equation 2, the parameters $E_0$, $D_0$ and $n$ only depend on the physical properties of QDs, and can be determined from the experimental data.

In order to determine these three parameters, we took the logarithm on both sides of

Equation 2.  Assuming a constant heating temperature $T_0$, we obtained

$$\ln\left|\Delta E_{gap}\left(T_0,t\right)\right| = n\ln t + A \qquad (3)$$

where $A$ is a constant.  **Equation 3** suggests that the logarithm of the energy shift, $\ln(\Delta E_{gap})$, is a linear function of the logarithm of the time $t$, $\ln t$, with the time component $n$ as the slope of the logarithmic relationship.

**Figure 4a** plots the time-dependent PL peak shift in logarithmic scale for the CdSe/ZnS quantum dots heated at 500 $^o$C.  The PL spectra were plotted in Figure 2a.  From a linear fitting, the time exponent $n = 0.21$ when the QDs were heated at 500 $^o$C. Similarly, we determined the time exponent $n$ from experimental PL data of the QDs heated at other heating temperatures $T$.  Figure 4b plots the determined $n$ at a series of heating temperatures $T$ from 150 $^o$C to 500 $^o$C in air.  The average value $n = 0.21$. It is interesting to note that for a typical mass diffusion process, $n = 0.5$.  However, for CdSe/ZnS core-shell quantum dots we found that $n$ was much lower, suggesting that the mass diffusion was affected by many parameters of the QDs.

Next, we assumed a constant heating time $t_0$:

$$\ln\left|\Delta E_{gap}\left(T,t_0\right)\right| = -\frac{E_b}{2\kappa_B T} + B \qquad (4)$$

where $B$ is a constant.  From **Equation 4** we found that the logarithm of the energy shift $\Delta E_{gap}$ scales inversely to the temperature $T$, with $E_b/(2\kappa_B T)$ as the slope.

**Figure 5** plots $\ln\Delta E_{gap} - 1/T$ curve for the CdSe/ZnS quantum dots when the heating time $t$ was 16 s.  The PL spectra were shown in Figure 3a.  From the fitting, $E_b = 0.24$ eV at heating time of 16 s.  We conducted a series of similar studies of the quantum dots at other heating time $t$ from 0.5 s to 128 s.  Figure 5b plots $E_b$ fitted from the $\ln(\Delta E_{gap}) - 1/T$ curves at other heating time $t$.  The energy barrier $E_b$ is about $0.20 - 0.30$ eV with an average value of 0.26 eV.

The diffusion factor $D_0$ was then calculated from the fitting values of constants $A$ and $B$. $D_0 = 2.57 \times 10^{20}$ m$^2$/s for the first kind of CdSe/ZnS QDs with original PL peak of 582 nm, falling within the reasonable range of self-diffusivity value. [20]

Similarly, other kinds of CdSe/ZnS QDs were also fast heated in air. Figure S5 shows PL spectra of CdSe/ZnS core-shelled QDs with original PL peak of 620 nm. The PL peak also blue-shifted with heating temperature $T$ (Figure S5a) and heating time $t$ (Figure S5b). The energy barrier of the mass diffusion $E_b$, time component $n$, and diffusion factor $D_0$ were then determined for the second kind of QDs (Figure S6), $E_b = 0.375$ eV, $n = 0.25$, $D_0 = 7.55 \times 10^{-20}$ m$^2$/s. Compared with $E_b = 0.26$ eV, $n = 0.21$, $D_0 = 2.57 \times 10^{-20}$ m$^2$/s of the first kind of QDs with original PL peak of 582 nm, $E_b$ of the second kind of QDs increased half and $D_0$ almost doubled. The differences should come from the size-dependent diffusion activation energy. According to the point defect mechanism, $D_0$ is proportional to $exp(E_b/(RT))$, where $E_b \propto exp\{-2S/[3R'(R/3h -1)]\}$ for nanospheres where $S$ is the bulk melting entropy, $h$ the atomic diameter, and $R'$ the ideal gas constant. $E_b$ is a weak function of $R$ while $D_0$ is much stronger. Therefor a slight change of the core size $R$ (such as 10 %) would cause a greater change of $E_b$ (50%) and more of $D_0$ (200%).

HRTEM image analyses also indicated that the diameter of the second kind of CdSe/ZnS QDs also distributed as a Gauss function before and after heating in air. The diameter of the QDs changed too little to conclude if the diameter $R$ changes or not after heating.

We had stored these heated QDs for six months in air at room temperature in dark and then measured PL spectra of these heated QDs. Same temperature- and time-dependent PL spectra were observed. The fact indicated that the QDs can permanently record thermal events and should be good thermal history nanosensors.

## 2.4. Thermal History Nano-sensors to Reconstruct Temperature History of Pulsed Heating Events

In most application cases, both temperature $T$ and evolution time $t$ are unknown factors in a individual thermal event while need to be recorded permanently and extracted later. A thermally induced band-gap change of QDs in a short time at a high temperature might be equivalent to a band-gap change in a long time at a low temperature. This non-uniqueness is a challenge for the thermal history retrieving. The lack of uniqueness could be compensated by using two kinds of, or even three or four kinds of QDs which have different response rates. The more kinds of QDs are applied, the more thermal history details could be recorded and retrieved.

Two kinds of QDs were employed here to permanently record thermal histories in order to extract both heating temperature $T$ and time $t$ of thermal events simultaneously. The temperature history (temperature $T$ and time $t$) of a heating event can be extracted from PL spectrum based on

$$\begin{cases} \left| \Delta E_{gap,1} \left(T,t\right) \right| = L_{0,1} \sqrt{D_{0,1}} e^{-\frac{E_{b,1}}{2\kappa_B T}} t^{n_1} \\ \left| \Delta E_{gap,2} \left(T,t\right) \right| = L_{0,2} \sqrt{D_{0,2}} e^{-\frac{E_{b,2}}{2\kappa_B T}} t^{n_2} \end{cases} \tag{5}$$

where 1 and 2 indicate the first and second kind of QDs respectively. The energy barrier of mass diffusion $E_b$, diffusion factor $D_0$ and time exponential $n$ for each kind of quantum dots were determined from PL spectra in Section 2.3.

Here we demonstrated how to use the two kinds of QDs as thermal history sensors to permanently record a heating history of a pulsed heating and reconstruct the thermal history later using Equation 5. We first deposited the two kinds of quantum dots on Pt hotwires. Here we chose the first kind of CdSe/ZnS QDs (original PL peak at 582 nm) with $E_{b,1} = 0.26$ eV, $n_1 = 0.21$, $D_{0,1} = 2.57 \times 10^{-20}$ m$^2$/s and second kind (original peak at 620 nm) with $E_{b,2} = 0.375$ eV, $n_2 = 0.25$, $D_{0,2} = 7.55 \times 10^{-20}$ m$^2$/s. Then the deposited QD mixture was fast heated at 400 $^o$C for 2 s in air. **Figure 6**a shows the heating history. The temperature $T$ was experimentally measured from the electrical resistance of the Pt hotwire and time $t$ was controlled by a Labview VI with a resolution of 70 milliseconds. The mixed QDs were heated up to 400 $^o$C in less than 140 ms from room temperature, heated at 400 $^o$C for 2 s, and then cooled down to room temperature in several milliseconds. The whole heating process was a typical fast heating without any temperature delay.

Then the PL spectrum of the mixed QDs was measured at room temperature after the fast heating and compared with that of unheated, as shown in Figure 6b. Both PL spectra were collected from a micron-scaled region where the excitation laser beam irradiated. We measured the PL spectra on different locations of the hotwire and obtained strong PL spectra corresponding to both kinds of quantum dots. There were two PL peaks before heating: the left peak coming from the first kind of CdSe/ZnS QDs with original PL peak of 582 nm and the right one from the second kind of QDs with original PL peak of 620 nm. After fast heating in air, both the PL peaks shifted to shorter wavelengths. The left peak shifted to 563 nm and the right one to 588 nm.

Using Equation 5, the heating temperature was solved as $T = 430 \pm 50$ °C and the heating time of $t = 1.9 \pm 0.2$ s. The reconstructed heating temperature $T$ and heating time $t$ were in agreement with the experimental data ($400 \pm 10$ °C and $2 \pm 0.07$ s).

In real chemical explosions or other heating events, the temperature of the heating event usually decreases very slowly with time $t$, not like the pulsed fast heating as shown here. Such heating processes are more complex than the fast heating events. Such practical heating events can be considered as two heating processes: one fast heating with heating temperature $T$ and heating time $t$ ($0 < t < t'$, where $t'$ is the whole fast heating time) and another heating delay expressed as $Te^{-\beta(t-t')}$ where $\beta$ is a delay factor. Such thermal histories can be permanently recorded and reconstructed from three kinds of core-shelled QDs. The related experiments and analysis are being carried out.

We also heated CdSe/ZnS QDs purchased from other companies in air. Similar PL shifts were observed. Additionally other kind of core-shelled semiconducting nanocrystals, such as CdTe/ZnS, were examined at different heating temperatures and durations. PL peaks also blue-shifted with increasing temperature and duration. Therefore the theoretical model can be applied to other core-shelled QDs as thermal history sensors. Besides PL spectroscopy, other optical techniques to determine band gap or microstructures, such as Raman spectroscopy and surface plasma resonance spectroscopy can also be employed to extract thermal histories recorded by core-shelled nanocrystals.

## 3. Conclusion

In summary, irreversible PL peak shifts were determined in core-shelled quantum dots CdSe/ZnS after thermal exposures because of quantum confinement effects. The characteristics of the PL peak shifts caused by the quantum confinement were utilized to permanently record thermal histories of heating events and retrieve thermal events later. Two kinds of CdSe/ZnS core-shelled QDs were employed to demonstrate recording of heating temperatures and heating durations of fast heating events simultaneously and reading out mechanism at microscale.

## 4. Experimental Section

*Cone-shelled CdSe/ZnS Quantum Dots.* The core-shelled CdS/ZnS QDs were suspended in toluene, purchased from Evident Technology Inc. The diameter of the CdSe/ZnS QDs was about several nanometers measured from HRTEM images. The thickness of ZnS shell was a few monolayers.

*Fast Heating.* The QDs were deposited on Pt hotwires. The Pt hotwires were heated up by a pulsed DC current that was controlled by a Labview VI. The DC current, electrical voltage on the hotwire, and the electrical resistance of the hotwires were measured by a Keithley 2400 source meter. The average temperature of the hotwires was calculated from the electrical resistance of the hotwires (Equation S1). The hotwires were heated up to a certain temperature $T$ (from 150 $^o$C to 500 $^o$C) in less than 140 milliseconds, then kept at a temperature for a certain heating time $t$ (from 0.5 seconds to 128 seconds). The hotwires were then cooled down to room temperature in less than 70 milliseconds. Compared with the heating duration of our pulsed heating events (several second), the ramp-up time and ramp-down time can be ignored and the whole heating events can be considered as ideal fast heating.

*PL Measurements.* The PL spectra were collected at room temperature with a WITec CRM200 confocal Raman system several days later after heating. An excitation laser beam with wavelength $\lambda$ = 532 nm was focused onto the QDs using a 100 × optical lens with numerical aperture $NA = 0.95$. The microscope objective produced a laser spot size of about 1 μm in diameter. The illumination and backscattering collection system consisted of a confocal microscope coupled to a single-grating spectrometer (a 1800 grooves/mm grating) and a thermoelectric-cooled charge-coupled-device cooled to -60 $^o$C. The spectra were collected with 1 s of integration time with a resolution of 1 cm$^{-1}$. The laser power was 15 mW and the laser intensity irradiated on QDs was manually adjusted. The laser intensity irradiated on QDs was kept as low as possible to avoid laser induced heating during the PL data acquisition.

*HRTEM Studying.* TEM samples of unheated QDs were prepared by dropping the QDs solutions on TEM grids, and left dry in ambient air. The fast-heated QDs on Pt hot-wires were directly observed on TEM. Both kinds of QDs were examined on a JEOL 2010F TEM.

**Supporting Information**

Supporting Information is available online from the Wiley Online Library or from the author.

**Acknowledgements**

This research is financially supported by Defense Threat Reduction Agency under grants HDTRA1-10-1-0001, HDTRA122221 and Small Grants Program from University of Houston. The authors thank Dr. Daniel Kraemer for helps on Labview programming and Dr. Dong Cai for 3D plotting.

**Figure 1.** Illustration of the energy band diagram of core-shelled CdSe/ZnS QDs before (a) and after (b) thermally inducing mass diffusions at interfaces between CdSe cores and ZnS shells. Red core: CdSe; Green shell: ZnS; Blue shell: layer of mass diffusion.

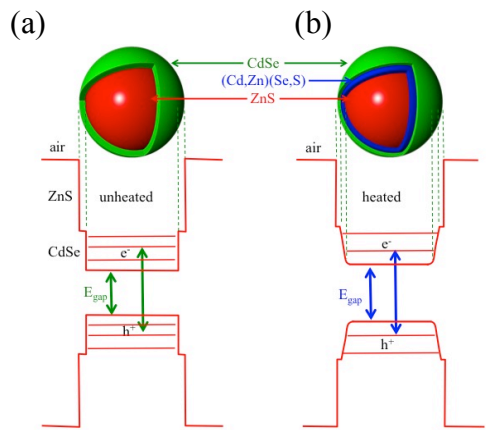

**Figure 2.** Temperature dependence of optical properties of the first kind of QDs. (a) PL spectrum of CdSe/ZnS QDs with original PL peak of 582 nm after being heated at different temperatures for 16 s. Each spectrum was recorded at room temperature after the heating events. The PL intensity was normalized for comparison. (b) Temperature dependence of PL peak wavelength of the heated QDs. The wavelength of each PL peak was fitted from PL spectra shown in (a).

(a)                                    (b)

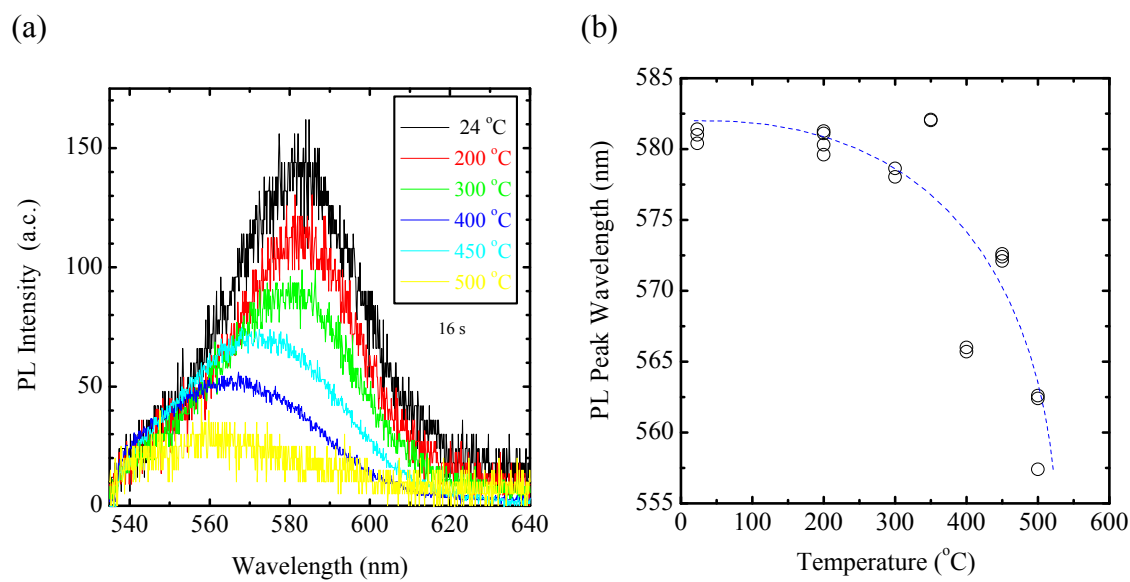

**Figure 3.** Time dependence of optical properties of the first kind of QDs. (a) PL spectra of CdSe/ZnS QDs heated at 500 °C. Each spectrum was measured at room temperature after the QDs were heated in air. (b) Dependence of PL peak wavelength on heating time $t$.

(a)

(b)

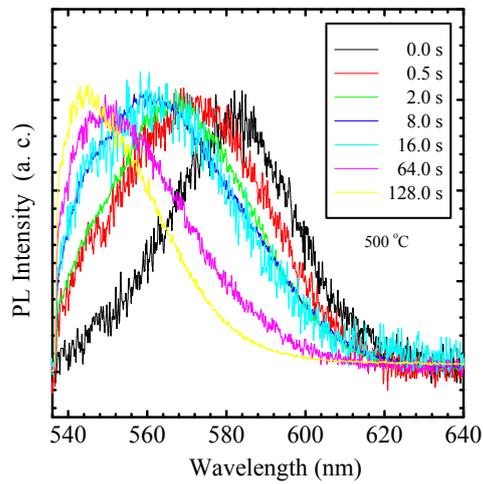
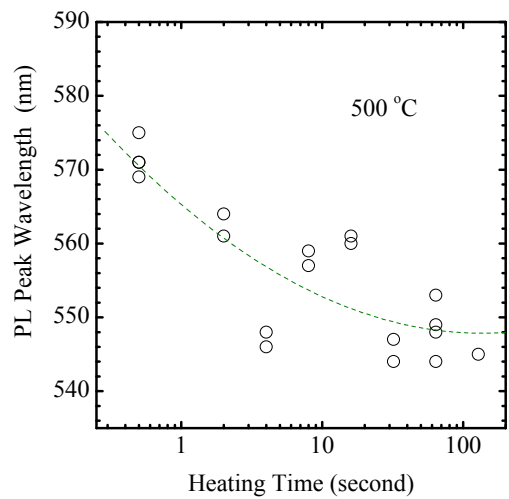

**Figure 4.** Time component *n* of the first kind of CdSe/ZnS QDs with original PL peak of 582 nm. (a) Blue-shift $\ln(\Delta E)$ with $\ln t$ for the CdSe/ZnS quantum dots. QDs were heated at 500 °C for different time *t* in air. (b) Fitted *n* at various heating temperatures from 300 °C for 500 °C. Each *n* was determined from related $\ln(\Delta E) - \ln t$ relationships.

(a)                                    (b)

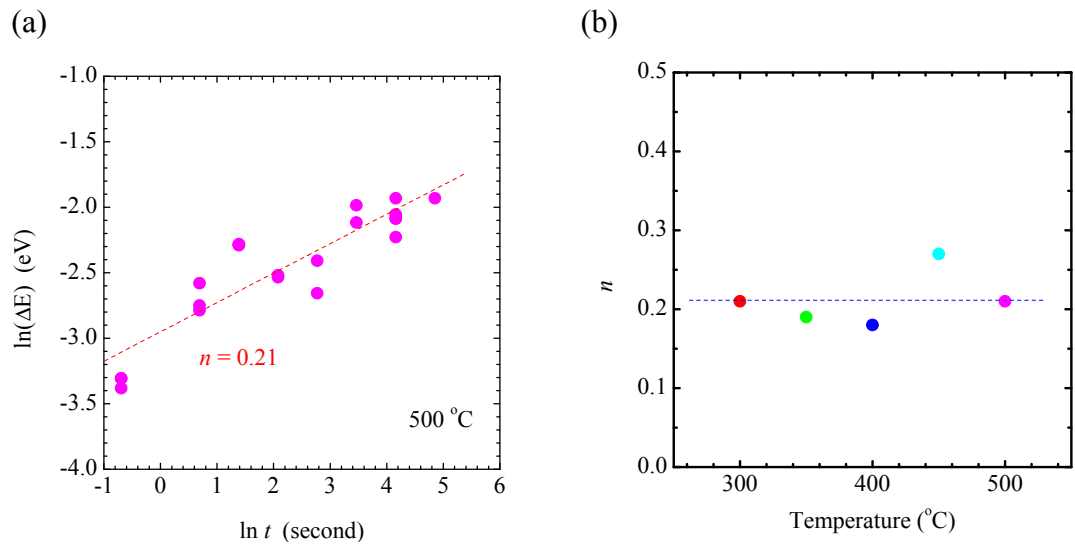

**Figure 5.** Diffusion barrier energy $E_b$ of the first kind of QDs with original PL peak of 582 nm. (a) Blue-shift $\ln\Delta E_{gap}$ with inverse temperature $1/T$. $t$ = 16 s. (b) Fitted $E_b$ at various heating time $t$ from 1 s to 128 s. $E_b$ for 16 s was determined from $\ln(\Delta E) - 1/T$ relationship shown in (a).

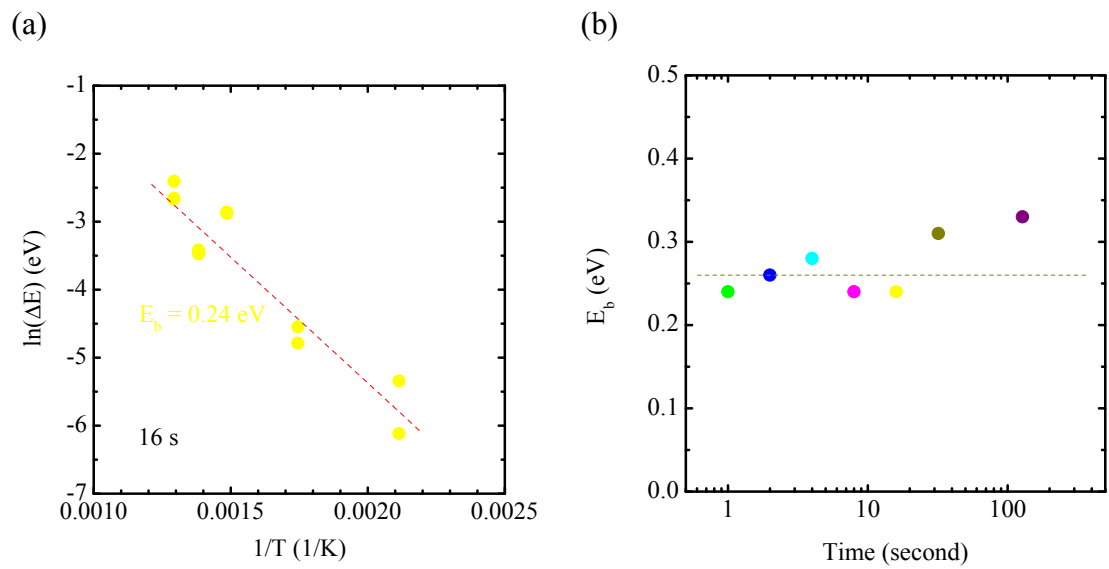

**Figure 6.** Thermal history of a fast heating event without time delay. (a) A fast-heating event without time delay. The temperature of the fast heating was calculated from electrical resistance of the hotwire based on Equation S2. The electrical resistance of the hotwire was recorded by a LabVIEW VI with a time resolution of 70 ms. (b) PL spectra of the mixed QDs. Blue circles: PL of unheated QDs. Red triangle: PL of heated QDs. Solid curves: fitted PL peak of QDs with peak of 620 nm. Dashed curves: fitted PL peak of QDs with peak of 582 nm. Blue: unheated QDs. Red: heated QDs.

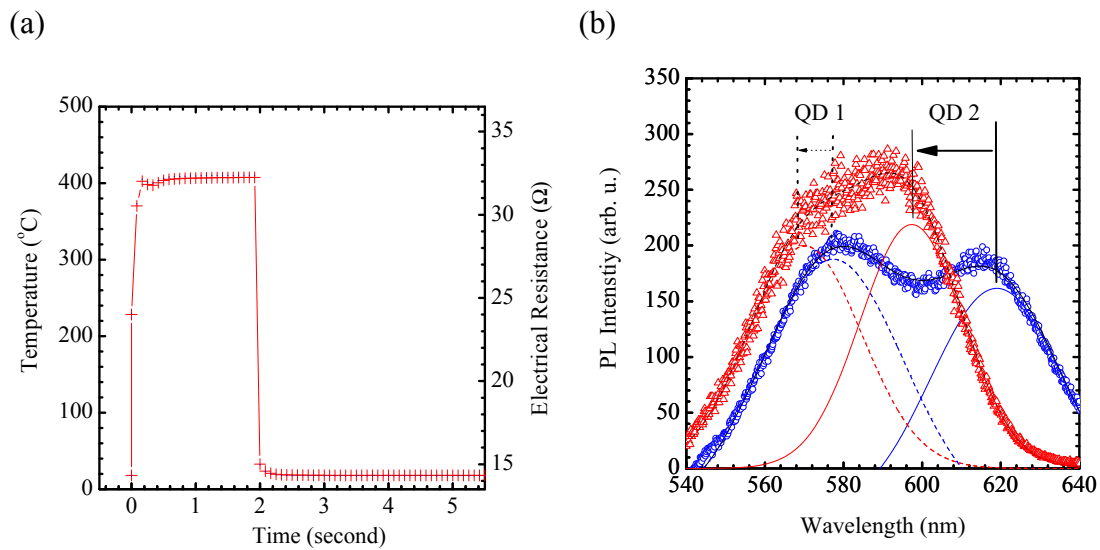

The table of contents entry should be 50−60 words long (max. 400 characters), and the first phrase should be bold.

**Thermal histories** are permanently recorded by CdSe/ZnS core-shelled quantum dots at microscale using photoluminescence spectroscopy and reconstructed later using a theoretical model. Theoretical equations are established between the photoluminescence shift and heating temperature / duration of thermal events. Here we demonstrate how to reconstruct the temperature and duration of a fast heating event simultaneously from photoluminescence spectra of the quantum dots.

Yucheng Lan,* Hui Wang, Nitin Skula, Xiaoyuan Chen, Yalin Lu, Gang Chen, and Zhifeng Ren*

**Determination of Thermal History by Photoluminescence Change of Core-Shelled Quantum Dots Going Through Heating Events**

ToC figure ((Please choose one size: 55 mm broad × 50 mm high **or** 110 mm broad × 20 mm high. Please do not use any other dimensions))

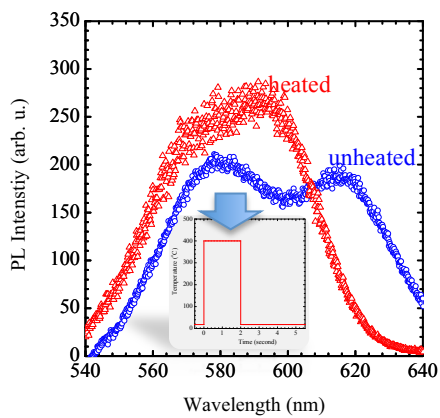